\documentclass{osa-article}

\journal{osajournal}

\articletype{Research Article}

\begin{document}

\title{Phononically shielded photonic-crystal mirror membranes for cavity quantum optomechanics}

\author{Georg Enzian,\authormark{1,*} Zihua Wang,\authormark{1} Anders Simonsen,\authormark{1} Jonas Mathiassen,\authormark{1} Toke Vibel,\authormark{1} Yeghishe Tsaturyan,\authormark{1,2} 
Alexander Tagantsev,\authormark{3} 
Albert Schliesser,\authormark{1} and Eugene S. Polzik\authormark{1}}

\address{\authormark{1}Niels Bohr Institute, University of Copenhagen, Blegdamsvej 17, 2100 Copenhagen, Denmark\\
\authormark{2} present address: Pritzker School of Molecular Engineering, University of Chicago, Chicago, IL 60637, USA\\
\authormark{3}Swiss Federal Institute of Technology (EPFL), School of Engineering, Institute of Materials Science, CH-1015 Lausanne, Switzerland
}

\email{\authormark{*}georg.enzian@nbi.ku.dk} 



\begin{abstract*}
%
We present a highly reflective, sub-wavelength-thick membrane resonator featuring high mechanical quality factor and discuss its applicability for cavity optomechanics. The $88.5~\text{nm}$ thin stoichiometric silicon-nitride membrane, designed and fabricated to combine 2D-photonic and phononic crystal patterns, reaches reflectivities up to 99.89~\% and a mechanical quality factor of $2.9 \times 10^7$ at room temperature. 
We construct a Fabry-Perot-type optical cavity, with the membrane forming one terminating mirror. 
The optical beam shape in cavity transmission shows a stark deviation from a simple Gaussian mode-shape, consistent with theoretical predictions.
We demonstrate optomechanical sideband cooling to mK-mode temperatures, starting from room temperature.
At higher intracavity powers we observe an optomechanically induced optical bistability.
The demonstrated device has potential to reach high cooperativities at low light levels desirable for e.g. optomechanical sensing and squeezing applications or fundamental studies in cavity quantum optomechanics, and meets the requirements for cooling to the quantum ground state of mechanical motion from room temperature.
\end{abstract*}
\section{Introduction}
High optomechanical cooperativity is instrumental for many optomechanical schemes and applications. It quantifies the coherent interaction between optical and mechanical degrees of freedom of a system in relation to their respective dissipation rates. It is most commonly defined as $C = 4 g^2/(\kappa\gamma_\textrm{m})$, where $g = g_0 |\alpha|$ is the optomechanical coupling rate, enhanced from the bare coupling rate $g_0$ by a coherent drive field of amplitude $|\alpha|$, $\kappa$ the optical intensity decay rate of the cavity, and $\gamma_\textrm{m}$ the mechanical energy decay rate \cite{aspelmeyer_cavity_2014}. 
In many cases cooperativity is a parameter representing the limit of device performance in sensing or optomechanical quantum state engineering schemes. Thus finding ways to increase it is desirable both for practical applications as well as fundamental studies in cavity optomechanics.

Reaching high cavity-optomechanical cooperativity typically entails confining light and mechanical vibration into a small region of space, while minimizing 
the dissipation. Coupling can arise due to, e.g., the radiation pressure exerted by the light onto mechanically compliant material boundaries, due to electrostriction in the bulk material, or due to electron-hole pair generation by cavity photons in semiconductor membranes \cite{usami_optical_2012}, and these mechanisms have been exploited to reach high optomechanical cooperativities \cite{groblacher_observation_2009, bahl_observation_2012}. 

Making use of the radiation-pressure coupling, thin dielectric membranes, suspended in or at the end of an optical cavity, 
have been an appealing system for cavity optomechanics \cite{jayich_dispersive_2008}.
Here, the mechanical and optical resonators can be designed relatively independently compared with other optomechanical platforms and optimized in order to mitigate losses. In particular, silicon nitride (SiN) membranes, attractive due to low optical absorption and high mechanical quality factors, which are a result of pre-tension in the membranes \cite{gonzalez_brownian_1998}, have been successfully used in a plethora of optomechanical experiments, e.g. \cite{wilson_cavity_2009, camerer_realization_2011, purdy_observation_2013, purdy_strong_2013, nielsen_multimode_2017, rossi_measurement-based_2018, brubaker_optomechanical_2022}.
Tsaturyan et al. \cite{tsaturyan_ultracoherent_2017} have produced \textit{Q-f} products above $10^{17}$ by patterning the membrane with a phononic-crystal structure, recently enabling  several demonstrations of quantum-limited operation and exploitation of quantum effects in optomechanics \cite{rossi_measurement-based_2018, mason_continuous_2019, chen_entanglement_2020, thomas_entanglement_2021, seis_ground_2022}.  

However, a bare SiN membrane will have limited reflectivity, consequently only weakly modifying the response of the optical cavity, in case of membrane-in-the-middle optomechanics, or not allowing for a significant optical cavity effect in the case of the canonical end-coupled resonator \cite{aspelmeyer_cavity_2014}.
It is not trivial to obtain high reflectivity with a dielectric slab or membrane which has a thickness less than the wavelength of light. Its reflection is limited to moderate values, with amplitude reflectivity reaching $r_d = (n^2-1)/(n^2+1)$ at the right thickness.
One way to make a light, mechanically compliant mirror for optomechanics was followed by Kleckner et al. \cite{kleckner_optomechanical_2011} who fabricated a layered dielectric mirror of microscopic lateral dimensions integrated with a SiN trampoline resonator.
A different approach to make even lighter and thinner mirrors is based on 2D photonic-crystal (PhC) slabs or 1D grating reflectors where Fano interference associated with guided resonances can cause very high reflectivities for near-perpendicular incidence \cite{fan_analysis_2002}.
The intriguing phenomenon of making a moderately reflecting dielectric slab a near perfect mirror by regular patterning has been experimentally demonstrated in numerous experiments, see e.g. \cite{peng_experimental_1996, kanskar_observation_1997, suh_displacement-sensitive_2003, lousse_angular_2004}. 

Several experimental efforts have been undertaken to explore PhC membranes as mechanically compliant elements for use in cavity optomechanics \cite{kemiktarak_mechanically_2012, bui_high-reflectivity_2012, kemiktarak_cavity_2012, makles_2d_2015, stambaugh_membrane---middle_2015}.
Norte et al. \cite{norte_mechanical_2016} presented PhC slabs integrated with an optomechanical trampoline resonator in an effort to make the mechanical quantum regime accessible from room temperature. 
Chen et al. \cite{chen_high-finesse_2017} demonstrated highly reflecting PhC slabs of stressed SiN and discussed their applicability for quantum optomechanics, with the main focus being the optical finesse. Here the resolved-sideband regime of cavity optomechanics was reached with the membrane serving as one end-mirror in an optical cavity.
Very recently, Zhou et al. \cite{zhou_cavity_2022} reported even higher reflectivity and optical bistability from radiation pressure.
However, the highest reflectivities have only been realized with relatively thick ($>200$~nm) and thus heavy as well as mechanically lossy \cite{villanueva_evidence_2014} membranes, and without the further shielding from mechanical dissipation provided by phononic-crystal patterning.

This work demonstrates a 88.5~nm thin SiN membrane featuring both photonic and phononic crystal patterns. 
The high mechanical quality factor is enabled by dissipation dilution and soft clamping \cite{tsaturyan_ultracoherent_2017}.
The increased reflectivity enables enhanced optomechanical coupling rates and cooperativities to be reached in membrane-in-the-middle or Fabry-Perot setups. Such enhanced cooperativity could enable improved squeezing and sensing capabilities, as well as ground state cooling starting from room temperature.
%
Regarding the structure of the article, section 2 describes the design, simulation, fabrication and characterization of the membrane. We determine the reflectivity as a function of wavelength, the mechanical quality factor and the effective mass. Section 3 presents measurements on a Fabry-Perot-type cavity with the membrane comprising one terminating mirror. We show that the beam-profile becomes highly non-Gaussian after transmission through the cavity, and verify that this agrees with simulations. We also find the membrane's reflection and parasitic loss as a function of cavity waist, and identify an optimum for our device.
Furthermore, as a demonstration of optomechanical performance, we cool the main mechanical membrane mode to $(550\pm30)$~mK from room temperature, demonstrating high optomechanical cooperativity.
Finally, we conclude with a discussion of our findings and an outlook on follow-up strands for future research.


\section{Highly reflective silicon nitride membranes featuring photonic and phononic patterning}

\subsection{Principle of operation and design}

\begin{figure}[htbp]
\centering
\includegraphics[width=0.85\textwidth]{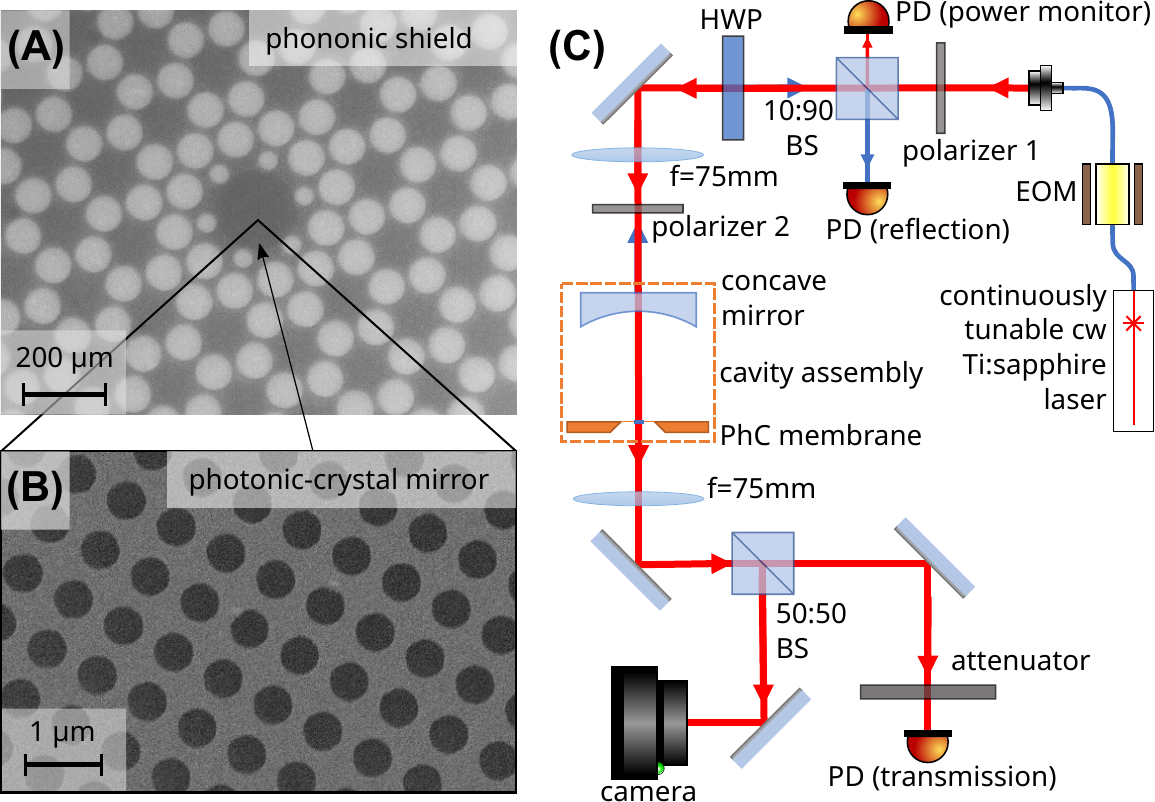}
\caption{Combined phononic-photonic-crystal patterned SiN membrane. 
(A) Optical microscope image of the membrane showing a hexagonal phononic pattern and central circular photonic-crystal section. (B) Close-up of the photonic-crystal pattern obtained with a scanning electron microscope. (C) Test setup for characterization of the optical properties of the membrane (BS: Beam splitter, EOM: Electro-optic modulator, HWP: Half-wave plate, PD: Photodetector, PhC: Photonic crystal). This setup was used for both plain reflectivity measurements without the concave mirror, as well as measurements on a cavity with the membrane as back-stop mirror.
\label{microscope}}
\end{figure}

We have designed and fabricated highly reflective PhC membranes from stoichiometric silicon nitride which, in addition to their PhC patterning, also feature a built-in phononic pattern, similar to the one reported in \cite{rossi_measurement-based_2018}. The membrane has a thickness of 88.5~nm, and has a central pad with a diameter of approximately 200~$\mu$m which constitutes a defect of the phononic crystal and houses the PhC pattern in its center. The photonic crystal is comprised of a square lattice of circular holes (see microscope images in Fig.~\ref{microscope} A-B) with lattice constant 733~nm, and circular hole diameter of 525~nm, and the pattern as a whole has a circular diameter of 200~$\mu$m.

\subsection{Fabrication}
\paragraph{Phononic patterning}
The phononic crystal membrane was fabricated following the procedure described in reference \cite{tsaturyan_ultracoherent_2017}.
Here we briefly summarize said fabrication process. We deposited 100~nm thick, stoichiometric silicon-nitride (Si$_3$N$_4$) using low-pressure chemical-vapor-deposition on a silicon wafer.
The phononic crystal pattern, as well as the square opening in the SiN film on the opposite side of the silicon substrate, are defined using UV lithography and transferred into the silicon nitride thin film via reactive ion etching. The wafer is stripped of any remaining photoresist and mounted in a PTFE (polytetrafluoroethylene) holder. While in the holder, the wafer is dipped into a buffered hydrofluoric acid (BHF) solution for 30 seconds, in order to remove the native oxide of the silicon substrate. Finally, the holder is submerged into a 30\% KOH solution at 80$^{\circ}$C, to etch through the silicon substrate and release the membrane structure. We finish the fabrication process by cleaning the wafer in a piranha solution.

\paragraph{Photonic patterning}
In order to add the PhC pattern, we insert an additional electron-beam lithography and etch step before the final release in KOH, using the same reactive ion etch recipe to transfer the photonic crystal pattern into the nitride. Based on separate testing of the etch, we expect the sidewall after etching to slope with $\approx 2^{\circ}$.

The final SiN thickness of 88.5~nm and refractive index of 1.997 were measured using ellipsometry, after completion of the entire fabrication process.
In addition, the thickness was measured independently by transferring a similar membrane (from the same wafer the experimentally relevant device originated from) onto a second carrier and then scanning an edge of the pattern with an atomic-force microscope (AFM) to measure the step-height.
The lattice constant was targeted at 734~nm, and determined to be (732.9 $\pm$ 3.3)~nm.
The hole diameter had a target value of 511.2~nm, and was measured to be (525.2 $\pm$ 6.8)~nm. These values were determined via analysis of images taken with a scanning-electron microscope (SEM), each including several holes of the pattern.

The experimentally measured parameters differ from the target parameters, which are the result of a numerical optimization in the design process (see more details in Appendix \ref{appendix_numerical}). 
Unfortunately, we found an unforeseen reduction of the nitride thickness by 13~nm, which is a large deviation compared with similar fabrication runs.
We believe it happened because our protective PTFE holder was not sealed properly, allowing BHF to leak in and attack the silicon nitride film hosting the photonic and phononic patterns.
The fabrication imperfections have shifted the reflectivity curve of the photonic mirror such that the highest reflectivity occurred around 833~nm instead of our target wavelength of 852~nm. For the realized parameters, we simulate a peak reflectivity of 99.986~\% for a normal-incident plane wave using finite element method (FEM) with Comsol Multiphysics.

\subsection{Wavelength dependence of the reflectivity}
\begin{figure}[htbp]
\centering
\includegraphics[width=0.95\textwidth]{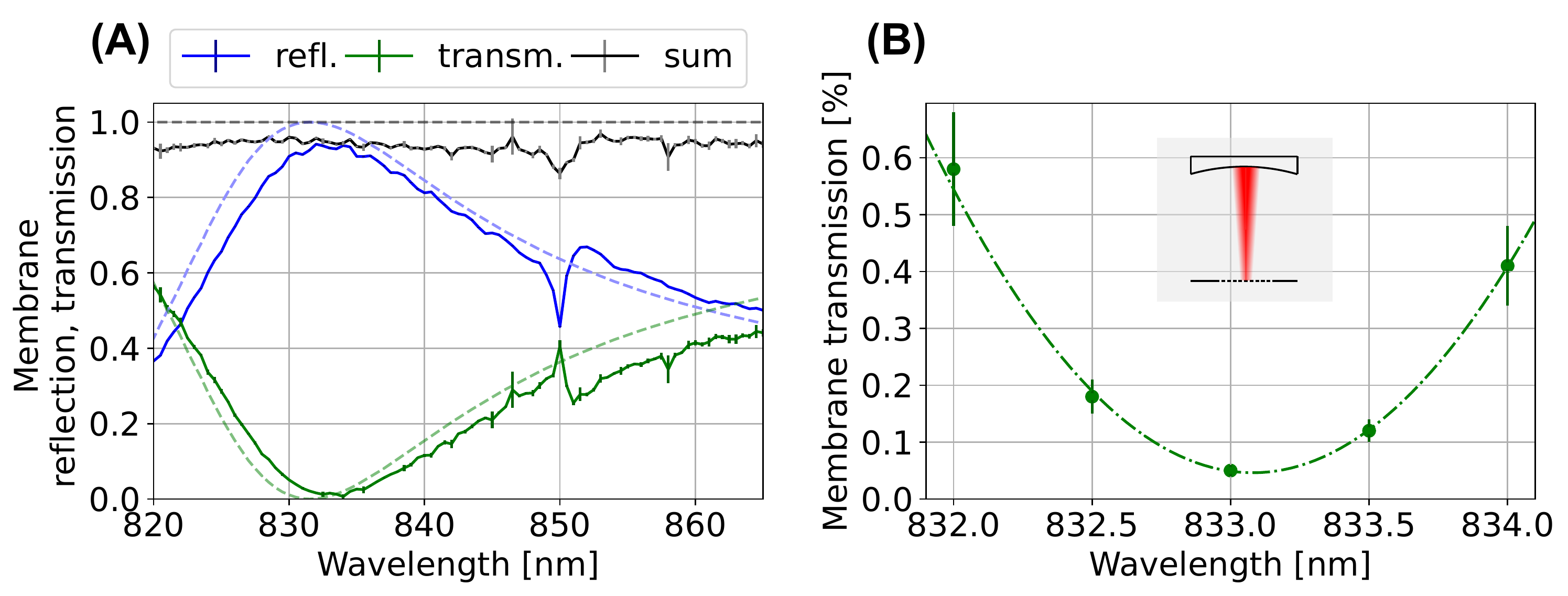}
\caption{Membrane reflectivity and transmission as a function of wavelength. (A) Measured by direct transmission and reflection measurements (no cavity) for a Gaussian beam with a waist of $w_0 = 16~\mu\textrm{m}$. The dashed line is the expected membrane reflectivity according to FEM simulation for a plane wave at normal incidence and simulation parameters given by the measured membrane parameters, assuming a PhC extending to infinity.
(B) Membrane transmission measured using a cavity with a mode waist $w_0 = 36~\mu\textrm{m}$, around peak reflectivity, with a quadratic function fit to the data (dash-dotted line).
\label{membraneRefl}
}
\end{figure}

We measured the reflectivity of the membrane as a function of wavelength between 820~nm and 865~nm, and for varying beam waists. To this end, we covered the range by tuning a continuously tunable continuous-wave Ti:Sapph laser with small increments and recording the reflection and transmission from the membrane at every point using the setup depicted in Fig.~\ref{microscope}C. 
These measurements carry a substantial uncertainty due to the difficulty in accurately measuring the small transmitted power, as well as lack of quick feedback on the quality of the alignment. We constructed an optical cavity, on which we report in section \ref{sec:cavityMeasurements}. Measurements on this cavity also allow us to reduce the uncertainty on the membrane's optical properties.
Fig.~\ref{membraneRefl} shows the power reflectivity and transmission of the membrane at a beam waist ($1/e^2$-intensity radius) of $w_0 = 16~\mu\textrm{m}$, as well as the transmission around the wavelength of peak reflectivity, obtained from a cavity measurement with a beam waist of $w_0 = 36~\mu\textrm{m}$.

We observe a reflection peak (transmission minimum) with a width of several nanometers around a center wavelength of 833.15~nm. The measurements resemble the general trend of the simulated reflection (transmission) performance (see Fig.~\ref{membraneRefl}A). The discrepancy is attributed to the use of a Gaussian incident beam compared to a plane incident wave in simulation, as well as uncertainties in the measured lattice constant and hole size. A sharp Fano-like feature is observed around 850~nm which is not reproduced by the plane-wave simulation. 


\subsection{Mechanical properties}
\subsubsection{Mechanical quality factor}\label{sec:mechQ}
\begin{figure}[htbp]
\centering
\includegraphics[width=0.97\textwidth]{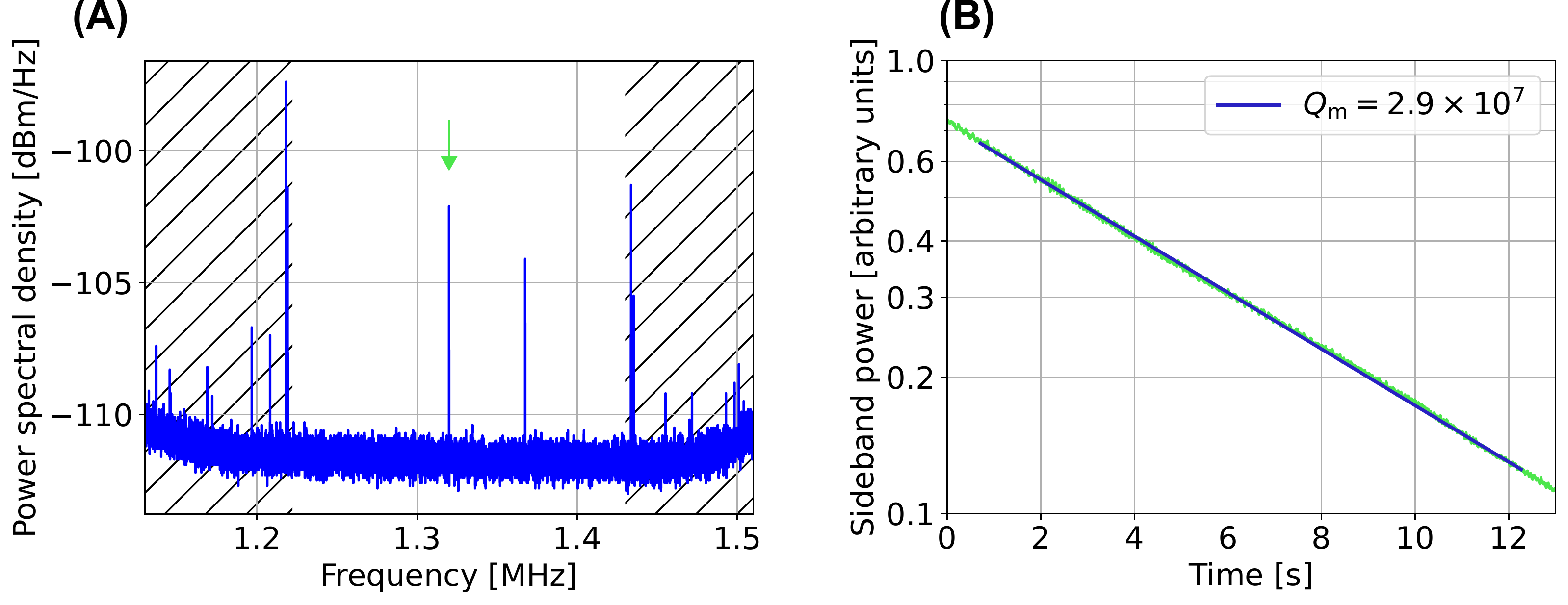}
\caption{Mechanical properties of the membrane, at room temperature. (A) Spectrum of thermal vibration of the membrane, measured using a homodyne interferometer, showing the mechanical band gap (unhatched) and defect modes lying inside it. (B) Ringdown measurement of the mechanical mode at $\Omega_\textrm{m} = 2\pi \times 1.32 \ \textrm{MHz}$ highlighted in A (green arrow).
\label{mechanicsFigure}}
\end{figure}

The mechanical quality factor of the membrane was measured by resonantly exciting the mechanical mode of interest with piezoelectric ring actuators mechanically coupled to the silicon frame of the membrane. Then, upon turning the excitation off, we observed ringdowns of the sideband amplitude with a Lock-In amplifier operating on the output of an optical homodyne interferometer. A typical thermal spectrum and a mechanical ringdown of a membrane are shown in Fig.~\ref{mechanicsFigure}. The lowest-order mechanical defect mode has a frequency of 1.32 MHz and a mechanical quality factor of $Q_\text{m} = 2.9 \times 10^7$, at room temperature.

\subsubsection{Effective mass}
In order to determine the effective mass\footnote{$m_\text{eff} = \int_{\mathrm{S}} dx dy \rho(x,y) h |z(x,y)/z(x_0,y_0)|^2$, where $\rho(x,y)$ is the membrane density (it is zero where we have etched the photonic and phononic pattern), $h$ is the membrane thickness, $z(x,y)$ is the out-of-plane displacement of the relevant mechanical mode, and the mode is normalized to the displacement at the readout point $(x_0,y_0)$. The integral is over the entire membrane surface $\mathrm{S}$.}, $m_\text{eff}$,
of the main mechanical mode of interest, we simulated the membrane's mechanical normal modes using FEM (Comsol Multiphysics). 
For this computation we did not add the PhC directly in the simulation as that would require a very dense numerical mesh and considerable computational resources. Instead, we approximated the crystal by scaling up its unit cell by a factor of 10 and reducing the number of holes accordingly. The resulting pattern has the same filling fraction as the fabricated membrane and well approximates its mechanical response as the unit cell is still kept much smaller than the mechanical wavelength.
The result is $m_\text{eff} = 6~\textrm{ng}$. Thus we infer the root-mean-square amplitude of mechanical zero-point fluctuations $x_\text{zpf} = (\hbar/(2 m_\text{eff} \Omega_\text{m}))^{1/2} = 1.0 \times 10^{-15}~\textrm{m}$.


\section{High-finesse cavity with PhC membrane as end mirror}
\label{sec:cavityMeasurements}
In order to assess peak reflectivity and parasitic loss of the membrane more precisely, we constructed an optical cavity with the membrane acting as the flat back-stop mirror. This configuration corresponds to the canonical toy model of cavity optomechanics in which the length of a simple Fabry-Perot-type optical resonator is modulated by the motion of one of its terminating mirrors which is assumed elastically suspended \cite{aspelmeyer_cavity_2014}. By introducing copper spacers into the cavity assembly we were able to adjust the cavity length and thus measure the membrane properties at a number of different beam waists. The beam waist is known to strongly affect the effective reflectivity obtained with PhC membrane reflectors \cite{moura_centimeter-scale_2018}. 
This dependence can be understood as a trade-off balance between increasing transmission due to clipping of the incident light field from the finite-sized PhC towards large incident mode waists, and increasing transmission towards narrow mode waists, due to the high angular composition of the light field, for which the crystal is less reflective.
At large mode waists the tails of the impinging light field will leak through the comparatively transparent edges of the PhC which consist of unpatterned SiN, resulting in increasing transmission.
At small mode waists, an impinging Gaussian mode, considered in the form of its plane-wave expansion, will have its plane-wave content probe the PhC on average under decreasingly orthogonal incidence, as the mode waist is reduced. Since the reflectivity is a decreasing function of incidence angle (see Fig.~\ref{clover}A), this effect results in an increase in transmission, too, leading to a reflectivity optimum between the two regimes.
As the in-coupling mirror of the cavity we used a plano-concave mirror with a radius of curvature of the high-reflection coated, concave surface of 25~mm, and a transmission of 1000~ppm around the wavelength of 830~nm.


\subsection{Cavity mode shape and clover-leaf shaped transmitted beam}\label{sec:cavity_mode_shape}
\begin{figure}[htbp]
\centering
\includegraphics[width=\textwidth]{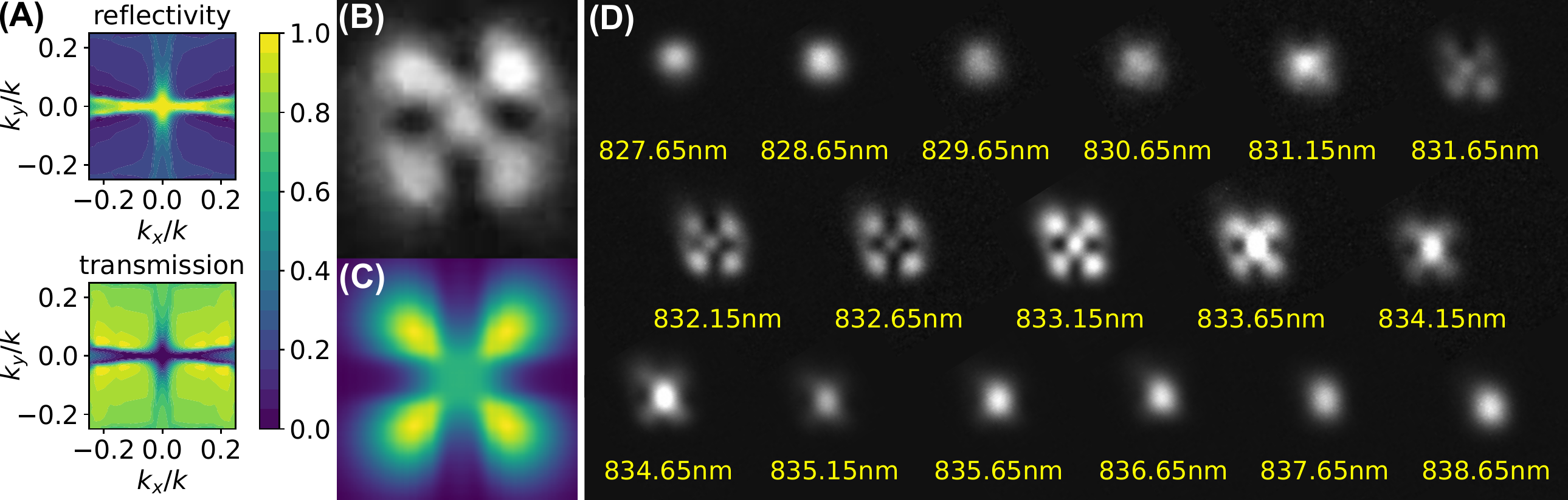}
\caption{Clover-leaf shaped beam shape after transmission through the cavity
due to angle-dependent reflectivity of the 2D PhC structure. 
(A) Total power reflectivity and transmission as a function of incidence angle for incident light polarized along x, showing near perfect reflection for perpendicular incidence.
(B) Experimentally observed clover-leaf shape at peak reflectivity. 
(C) Simulated transmission based on a Fox-Li cavity simulation and subsequent transmission through the membrane.
(D) Beam shapes of transmitted fields at wavelengths in the neighborhood of peak reflectivity, at a beam waist of $18.5~\mu\textrm{m}$.
\label{clover}}
\end{figure}

We observed that the field transmitted through the cavity has a clover-leaf-shaped intensity distribution, as depicted in Fig.~\ref{clover} for a beam waist of $18.5~\mu\textrm{m}$. 
This transmission pattern has not been reported in the previous work that used PhC mirrors in nitride \cite{bui_high-reflectivity_2012, makles_2d_2015, stambaugh_membrane---middle_2015, norte_mechanical_2016, chen_high-finesse_2017, zhou_cavity_2022}.
In order to investigate the origin of this shape, we set up a vectorial Fox-Li algorithm \cite{fox_resonant_1961, asoubar_simulation_2016}. 
The Fox-Li method calculates the fundamental mode of the cavity by iteratively propagating the complex electric field, represented on a 2D equidistant grid over the transverse plane, through the cavity round-trip. Free propagation of the field is handled by the FFT method under paraxial approximation. We extend the originally scalar method to a vectorial form, in order to account for the polarisation-dependent reflectivity of the PhC membrane.  
The algorithm takes as input the polarization- and angle-dependent complex reflectivity and transmission of the PhC membrane which we obtained from FEM simulation using Comsol Multiphysics 5.6 (see Fig.~\ref{clover}A and appendix \ref{appendix_numerical}). 
The Fox-Li simulation shows that the fundamental cavity eigenmode maintains a Gaussian beam profile for all explored cavity lengths, while the angle-dependent transmission through the PhC that makes light leak out of the rear of the cavity gives rise to a clover-leaf-like beam shape in the far field behind the cavity (see Fig.~\ref{clover}C).

%


\subsection{Determining membrane reflectivity and loss}
By measuring cavity reflection, transmission and linewidth at a number of different wavelengths around peak reflectivity, we determine the membrane's reflectivity and the parasitic loss introduced by the membrane. The parasitic loss is likely to be dominated by scattering loss, as SiN possesses very low optical absorption \cite{wilson_cavity_2009}. However the etching process might have altered material absorption and led to additional loss.
At this stage, without further analysis of local heat dissipation, we are not able to tell apart absorption from scattering. Thus we can only assess the lumped loss here. 
However, we note that the laser cooling measurements presented in section \ref{sec:sbcooling} did not show any signs of light-induced heating. The membrane's transmission is plotted as a function of wavelength in Fig.~\ref{membraneRefl}.
%

\subsubsection{Membrane reflectivity and parasitic loss estimation using resonance-pairing method}
Here we present a method that allows us to accurately determine the reflectivity $R_\textrm{mem}$, transmission $T_\textrm{mem}$ and parasitic loss $\beta$ of the membrane. 
To begin with, we state that for a high-finesse cavity the resonant cavity transmission $T$ is given by
\begin{equation}\label{eq:cavTransm}
    T = \frac{4 T_\text{mirr} T_\text{mem}}{\rho^2} \ ,
\end{equation}
where $T_\text{mirr}$ is the concave input mirror's transmission and $\rho = T_\text{mem} + T_\text{mirr} + \beta$ the round-trip loss.
Eq.~(\ref{eq:cavTransm}) already allows us to determine $T_\text{mem}$ from measurement of $T$ and $\rho$, given a single resonance and known input-mirror transmission. However, this approach requires calibration of the transmission, and is thus prone to systematic errors.
Therefore, instead we found a method for measuring membrane transmission and parasitic loss without the need for calibration of optical powers. The method relies on measuring resonant cavity transmission $T$ and finesse $\mathcal{F}$ at a small number (six, in our case) of different longitudinal mode resonances in a narrow wavelength range ($\approx$~1~nm).

It is furthermore assumed that the transmissivity of the input mirror $T_\textrm{mirr}$ is constant over this small wavelength range, and that chromatic aberrations of the incoupling optics are negligible. Negligible change across the wavelength range is also assumed for the mode matching of the input beam into the cavity, as well as the intra-cavity parasitic loss $\beta$, which can be considered as entirely due to the membrane. The last two assumptions are well founded in the observation that the mode shape of the fundamental cavity eigenmode remains Gaussian at all beam waists, according to our Fox-Li simulation, see section \ref{sec:cavity_mode_shape}.
We see that $\rho - T_\text{mem}$ is conserved when probing different resonances within a small wavelength range (due to the conservation of $\beta + T_\text{mirr}$). For any pair of resonances we can then write
\begin{equation}\label{eq:pair}
    \rho_1 - \rho_2 = T_\text{mem,1} - T_\text{mem,2} \ , 
\end{equation}
where $1,2$ represent measurement at two different resonant wavelengths $\lambda_1$ and $\lambda_2$, respectively.
If we consider two instances of Eq.~(\ref{eq:cavTransm}) at two different wavelengths, we can eliminate the (constant) coupling mirror transmission $T_\text{mirr}$ from them. Subsequently, we can substitute for the membrane transmission at $\lambda_1$ in Eq.~(\ref{eq:pair}), and obtain an expression for the membrane transmission at $\lambda_2$ as
\begin{equation}\label{Tmem2}
    T_\text{mem,2} = \frac{T_2 \rho_2^2}{T_1 \rho_1^2 - T_2 \rho_2^2} \left(\rho_1 - \rho_2\right) \ .
\end{equation}
Here $T_1, T_2$ and $\rho_1, \rho_2$ are the measured resonant cavity transmissions and round-trip losses at the two different resonant wavelengths $\lambda_1$ and $\lambda_2$, respectively.
By inserting Eq.~(\ref{Tmem2}) back into the definition of the round-trip loss, we find that the parasitic loss at $\lambda_2$ can be expressed as
\begin{equation}
    \beta = \rho_1 \rho_2 \frac{T_1 \rho_1 - T_2 \rho_2}{T_1 \rho_1^2 - T_2 \rho_2^2}  \ - \ T_\text{mirr} \  ,
\end{equation}
which does not require a calibration of the cavity transmission $T$.

\subsubsection{Results}
\begin{figure}[htbp]
\centering
\includegraphics[width=0.7\textwidth]{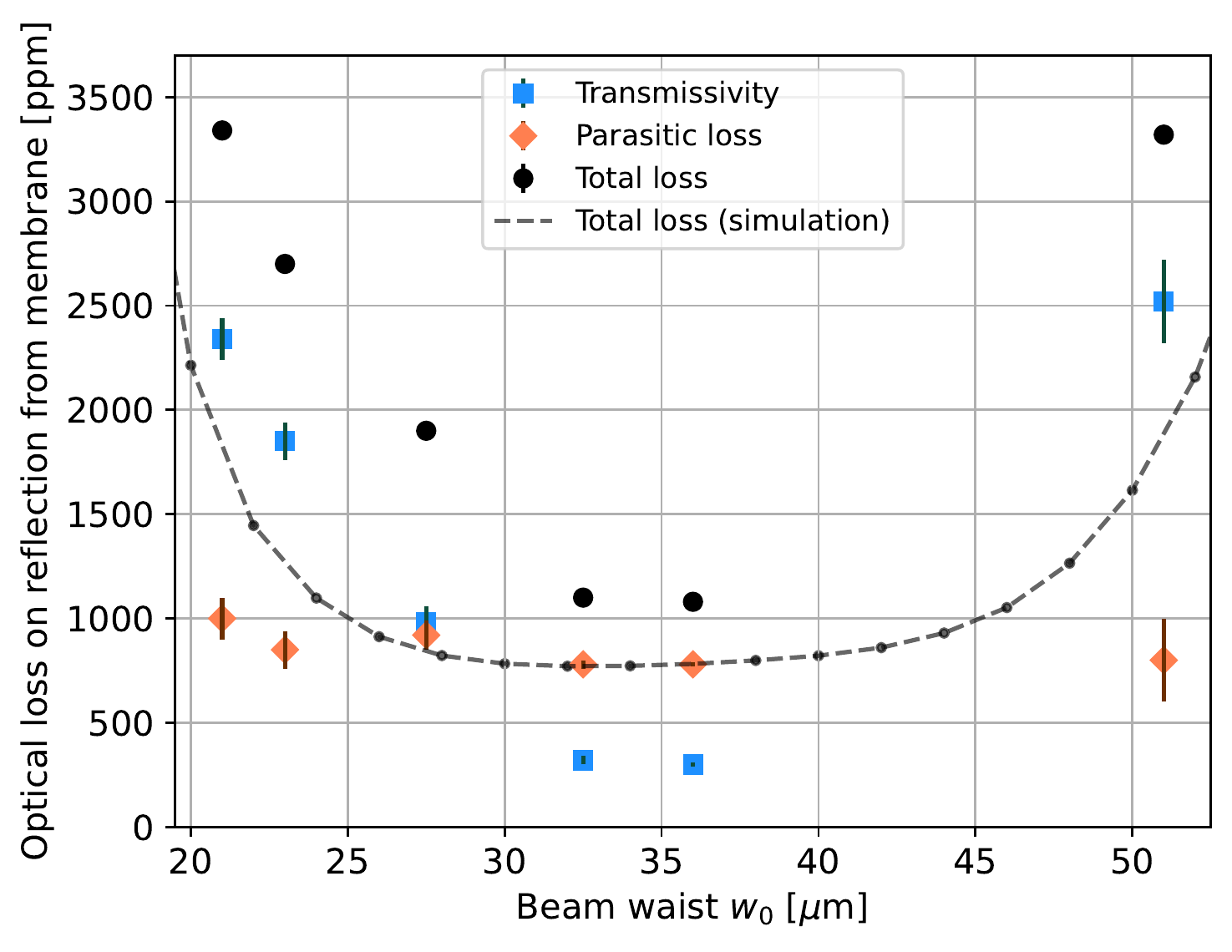}
\caption{Optical loss on the reflection from the membrane due to transmission and parasitic losses when illuminated at its peak-reflectivity wavelength of 833.15 nm, as obtained from cavity measurements at different cavity lengths. 
The dashed line corresponds to the total membrane-associated loss obtained from simulation. 
\label{loss_vs_waist}}
\end{figure}

We measured the effective transmission and loss of the membrane at a number of different Gaussian mode waists, which are accessed through changes to the cavity length with a concave in-coupling mirror of constant radius of curvature (compare Fig.~\ref{microscope}). 
The results are shown in Fig.~\ref{loss_vs_waist}.
The cavity finesse reaches values as high as $\mathcal{F} =2880\pm10$ at a nominal Gaussian mode waist of $w_0 = 36~\mu\textrm{m}$ (cavity length 6.34 mm). That corresponds to a cavity loss through the membrane of (1080 $\pm$ 10)~ppm and membrane reflectivity of 99.89~\%, or mirror membrane finesse of $\mathcal{F}_\text{mem}=5820\pm50$. The parasitic loss induced by the membrane is determined to be (780 $\pm$ 10)~ppm, and seems to be largely independent of the mode waist. The membrane transmission reaches minimal values of (300 $\pm$ 10)~ppm.

The simulated total loss is almost entirely due to transmission. While the simulated total loss captures the overall trend in the experimental data, the measured transmission loss is even lower than simulated in the optimal range around 35~$\mu$m mode waists. The parasitic loss is constant across different beam waists and is close to the value reported in \cite{chen_high-finesse_2017}, possibly hinting towards a shared limitation. Very recent work using a hexagonally patterned PhC membrane observed even lower parasitic loss and transmission of less than $170~\mathrm{ppm}$ \cite{zhou_cavity_2022}.


\subsection{Optical bistability}\label{bistab}
\begin{figure}[htbp]
\centering
\includegraphics[width=0.4\textwidth]{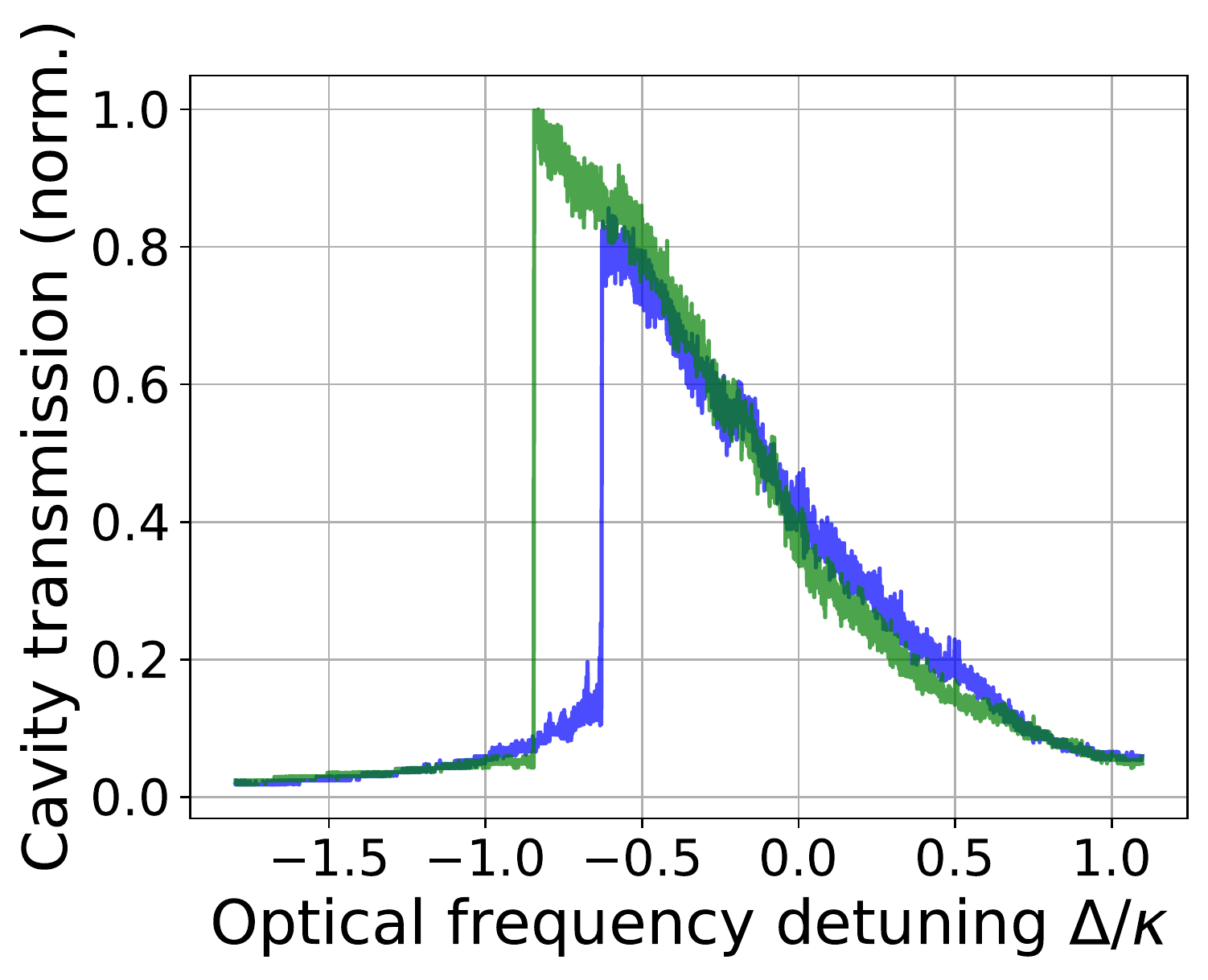}
\caption{Observed hysteresis of an optical resonance of the cavity due to the radiation-pressure-induced optical bistability. Cavity transmission as a function of optical frequency detuning from the empty-cavity eigenfrequency, when scanning the laser from red to blue (blue curve), and blue to red (green curve).
\label{hysteresis}}
\end{figure}

The radiation pressure force in a cavity optomechanical system can lead to an optical bistability \cite{dorsel_optical_1983, aspelmeyer_cavity_2014, zhou_cavity_2022}. We measure the associated hysteresis when sweeping the cavity length around the resonance condition with a coherent drive field, such that the detuning changes from red to blue and blue to red, respectively, see Fig.~\ref{hysteresis}. 
The static spring constant $k_\text{static}$ of out-of-plane displacement of the entire membrane obtained via FEM simulation is 34 N/m (compared with an effective spring constant of $k_\text{eff}$=410 N/m for the main mechanical defect mode).
For red detunings, the bistability limits the experimentally usable intracavity drive power to $P_\text{thr} = \sqrt{3} k_\text{static} c \lambda/(9 \mathcal{F})$, see e.g. \cite{zhou_cavity_2022}. For our system, the maximum intracavity power with peak membrane reflectivity is given by $P_\text{cav}^\text{max}$~=~0.57~W.
For the measurement shown in Fig.~\ref{hysteresis} the finesse was 2800, corresponding to a bistability threshold of 0.58 W. The peak intracavity power was 2.0 W. 


\subsection{Optomechanical sideband cooling}\label{sec:sbcooling}
\begin{figure}[htbp]
\centering
\includegraphics[width=0.7\textwidth]{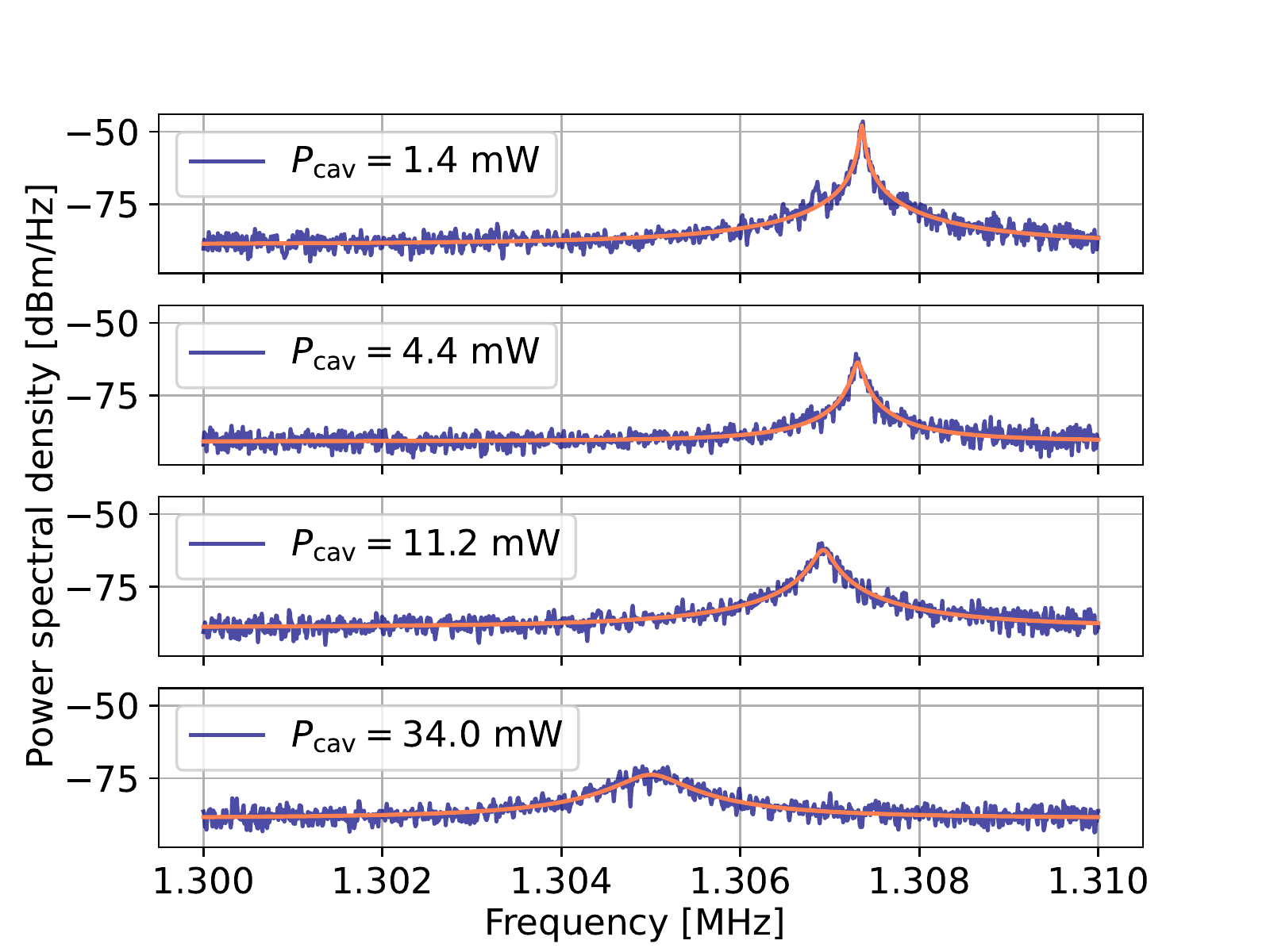}
\caption{Optomechanical sideband cooling of the main membrane mode by dynamical back-action. Different attenuators were used in order to not saturate the transmission APD photodetector. The dynamical broadening of the mechanical motion reaches $(616~\pm~30)$~Hz.
\label{om_sb_cooling}}
\end{figure}

We drive the optomechanical cavity with the laser red-detuned by one half of the cavity linewidth, which was at 13~MHz for this particular experiment. The detuning is stabilized via a feedback loop using the cavity transmission and controlling the cavity length via a piezoelectric actuator. We record spectra of the intensity fluctuations using the transmission photodetector. The observed spectra are shown in Fig.~\ref{om_sb_cooling}. 
We observe cooling of the thermal mechanical motion of the membrane via the dynamical back-action of the light field on the membrane. For the highest recorded power we observe a dynamically broadened mechanical linewidth of $\Gamma_\text{tot}/2\pi = (616\pm30)$~Hz.
From this we infer a mode temperature of (550$\pm$30)~mK. It should be noted that for the cooling measurements presented, due to suboptimal vacuum conditions in the experimental chamber, the membrane experienced considerable gas damping, only operating at $Q_\text{m} = 1.14 \times 10^6$. Under vacuum conditions as present in our interferometer chamber (compare section \ref{sec:mechQ}), a mode temperature of 22~mK would be observed at the same levels of cooling power. 
As the intracavity optical power is increased, we do not see the area under the thermal peak increase beyond the theoretical prediction, which indicates the absence of light-induced heating.


\subsection{Achievable cooperativity and potential for ground state cooling from room temperature}
Finally, let us consider the cooperativity achievable with our membrane when operating the cavity in the regime of linearized optomechanics with a coherent drive field. 
While membrane setups involving multiple sub-cavities, like e.g. the so-called membrane-at-the-edge setup \cite{dumont_flexure-tuned_2019}, can offer advantages w.r.t. cooperativity, we will consider the simple Fabry-Perot setup here, which we have also implemented experimentally. 
The single-photon cooperativity $C_0$ gets enhanced by the mean number of photons in the cavity $\bar{n}_\text{cav} = 2 L \lambda P_\text{cav}/(h c^2)$, to reach a maximum value of
\begin{equation}
    C^\text{max} = C_{0} \bar{n}_\text{cav}^\text{max} = \frac{32 \pi \omega_c x_\text{zpf}^2 P_\text{cav}}{h c^2 (T_\text{mem}^\text{min} + \beta) \gamma_m} \ , 
\end{equation}
for the cooperativity. Here the intracavity power $P_\text{cav}$ is limited due to the optical bistability. We note that the achievable cooperativity is independent of the cavity length $L$ for a given intracavity power $P_\text{cav}$ as it is limited by finesse related to parasitic membrane losses.

These numbers correspond to a cooperativity per intracavity power of $C/P_\text{cav}=1.2\times10^7~\text{W}^{-1}$ with the membrane at room temperature, compared with e.g. $C/P_\text{cav}=8.2\times10^5~\text{W}^{-1}$ in recent related work \cite{zhou_cavity_2022}. 
If the cavity was operated near the intracavity power corresponding to the bistability threshold of $P_\text{cav} = 0.57~\text{W}$ the cooperativity would reach $C = 6.8 \times 10^6$, and thus  exceed the thermal occupation of the mechanical mode $\bar{n}_\text{th} = 4.6 \times 10^6$ at room temperature.
Thus the membrane demonstrated here will allow optomechanical sideband cooling to the quantum ground state from room temperature in a sideband resolved cavity, which is feasible with the membrane demonstrated in this work (minimum cavity length of 5~cm). 
Achieving this in practice requires sufficiently low levels of technical noises, such as laser phase noise and mirror thermal noise \cite{saarinen_laser_2022}.
In fact, high-quality mechanical mirror membranes as demonstrated in this work could be combined with focusing photonic-crystal patterning \cite{guo_integrated_2017} in order to reach very low (mirror) thermal noise and facilitate cavity quantum optomechanics at room temperature.


\section{Conclusion and Outlook}
Our work demonstrates a thin (88.5~nm) optomechanical membrane with both enhanced reflectivity and high mechanical quality factor reached by use of a combination of photonic-crystal and phononic-crystal patterning of the membrane. These properties allow for high optomechanical cooperativities to be reached by employing the membrane as the end mirror in a Fabry-Perot-type resonator.
The membrane being thin allows for large mechanical quality factors to be reached, as well as keeps the effective mode mass low, aspects which increase the quantum prospects of the platform. Furthermore the membrane can be employed as a highly reflective ultra-low-noise mirror, using the absence of mechanical modes in the phononic-bandgap regions of the spectrum.

Having a look at further perspectives,
in order to reach large single-photon optomechanical cooperativity, the cavity can be made short. With both the bare coupling rate $g_0$ and the cavity linewidth $\kappa$ scaling as $1/L$, the cooperativity $\sim g_0^2/\kappa$ scales as $1/L$, too.
For a cavity length of $2~\mu\text{m}$, which seems technically feasible \cite{dumont_flexure-tuned_2019}, a single-photon cooperativity of $C_0 \sim 200$ can be expected, at room temperature. The room-temperature availability of multiple mechanical modes deep in the high single-photon cooperativity regime might enable implementation of nonlinear optomechanical schemes for non-classical state generation \cite{borkje_critical_2020}.
If down-scaling of the mechanical frequency is desired, e.g. for exploration of quantum squeezing by fast measurement \cite{meng_mechanical_2020}, large membranes are required. In this direction, Moura et al. have demonstrated centimeter-scale photonic-crystal mirror membranes \cite{moura_centimeter-scale_2018}. 
During the preparation of this manuscript, we became aware of work by Zhou et al. \cite{zhou_cavity_2022} demonstrating photonic-crystal membranes featuring hexagonal patterning reaching reflectivities above 0.9998, and also observing the optical bistability. 
At cryogenic temperatures the mechanical quality factors of membrane resonators mechanically similar to ours have been shown to reach exquisitely high values of the order $10^9$ \cite{tsaturyan_ultracoherent_2017}, highlighting a path towards further increase of the cooperativity. 
%


\appendix

\section{Appendix: Numerical simulation of photonic mirror}\label{appendix_numerical}
Numerical simulations of the photonic mirrors were performed using FEM implemented in Comsol Multiphysics (version 5.3 and 5.6). The simulation model considers a single unit-cell with periodic boundaries, thus modelling an infinite array. The software solves for the amplitude reflection and transmission of a plane-wave impinging on the structure. 

The optimal parameters for the lattice constant (734~nm), and the hole diameter (511.2~nm), were determined for a plane-wave at normal incidence with linearly polarized light and a fixed thickness (101.69~nm), refractive index (1.9861), and wavelength (852~nm). The thickness and refractive index were obtained from ellipsometry measurements on the SiN layer before the start of processing. Two parameters were tuned to maximize the reflection: the size of the holes and the lattice constant. The mesh was automatically generated by the software, but the mesh element size was limited to the optical wavelength divided by 7.

To generate the plots in Fig.~\ref{clover}A, the incidence angle of the plane-wave was varied while extracting the reflection and transmission. The input polarization was set linear and along the same direction for all incident angles, and the simulation output contains reflection and transmission for both the input polarization and the one orthogonal to it.



\begin{backmatter}
\bmsection{Funding}
This research has been funded by European Research Council (ERC) Advanced Grant QUANTUM-N under the EU Horizon 2020 (grant agreement no. 787520), Consolidator grant PHOQS (grant agreement no. 101002179), by Villum Fonden under a Villum Investigator Grant, grant no. 25880, and by the Novo Nordisk Foundation (grant NNF20OC0059939 ‘Quantum for Life’ and grant NNF20OC0061866), as well as the Independent Research Fund Denmark (grant 1026-00345B `ROOT-PHOQUS').
G.E. acknowledges support from the European Union's Horizon 2020 research and innovation programme under the Marie Sklodowska-Curie grant agreement no. 847523.

\bmsection{Acknowledgments}
The authors gratefully acknowledge helpful discussions with Jörg Helge Müller. 

\bmsection{Disclosures}
The authors declare no conflicts of interest.

\bmsection{Data availability} Data underlying the results presented in this paper are not publicly available at this time but may be obtained from the authors upon reasonable request.


\end{backmatter}




\bibliography{__Highly_Refl_Membranes_paper}

\begin{thebibliography}{10}
\newcommand{\enquote}[1]{``#1''}

\bibitem{aspelmeyer_cavity_2014}
M.~Aspelmeyer, T.~J. Kippenberg, and F.~Marquardt, \enquote{Cavity
  {Optomechanics},} {\protect\JournalTitle{Reviews of Modern Physics}}
  \textbf{86}, 1391--1452 (2014). ArXiv: 1303.0733.

\bibitem{usami_optical_2012}
K.~Usami, A.~Naesby, T.~Bagci, B.~Melholt~Nielsen, J.~Liu, S.~Stobbe,
  P.~Lodahl, and E.~S. Polzik, \enquote{Optical cavity cooling of mechanical
  modes of a semiconductor nanomembrane,} {\protect\JournalTitle{Nature
  Physics}} \textbf{8}, 168--172 (2012).

\bibitem{groblacher_observation_2009}
S.~Gröblacher, K.~Hammerer, M.~R. Vanner, and M.~Aspelmeyer,
  \enquote{Observation of strong coupling between a micromechanical resonator
  and an optical cavity field,} {\protect\JournalTitle{Nature}} \textbf{460},
  724--727 (2009).

\bibitem{bahl_observation_2012}
G.~Bahl, M.~Tomes, F.~Marquardt, and T.~Carmon, \enquote{Observation of
  spontaneous {Brillouin} cooling,} {\protect\JournalTitle{Nature Physics}}
  \textbf{8}, 203--207 (2012).

\bibitem{jayich_dispersive_2008}
A.~M. Jayich, J.~C. Sankey, B.~M. Zwickl, C.~Yang, J.~D. Thompson, S.~M.
  Girvin, A.~A. Clerk, F.~Marquardt, and J.~G.~E. Harris, \enquote{Dispersive
  optomechanics: a membrane inside a cavity,} {\protect\JournalTitle{New
  Journal of Physics}} \textbf{10}, 095008 (2008).

\bibitem{gonzalez_brownian_1998}
G.~I. González and P.~R. Saulson, \enquote{Brownian motion of a mass suspended
  by an anelastic wire,} {\protect\JournalTitle{The Journal of the Acoustical
  Society of America}} \textbf{96}, 207 (1998).

\bibitem{wilson_cavity_2009}
D.~J. Wilson, C.~A. Regal, S.~B. Papp, and H.~J. Kimble, \enquote{Cavity
  {Optomechanics} with {Stoichiometric} {SiN} {Films},}
  {\protect\JournalTitle{Physical Review Letters}} \textbf{103}, 207204 (2009).

\bibitem{camerer_realization_2011}
S.~Camerer, M.~Korppi, A.~Jöckel, D.~Hunger, T.~W. Hänsch, and P.~Treutlein,
  \enquote{Realization of an {Optomechanical} {Interface} {Between} {Ultracold}
  {Atoms} and a {Membrane},} {\protect\JournalTitle{Physical Review Letters}}
  \textbf{107}, 223001 (2011).

\bibitem{purdy_observation_2013}
T.~P. Purdy, R.~W. Peterson, and C.~A. Regal, \enquote{Observation of
  {Radiation} {Pressure} {Shot} {Noise} on a {Macroscopic} {Object},}
  {\protect\JournalTitle{Science}} \textbf{339}, 801--804 (2013).

\bibitem{purdy_strong_2013}
T.~P. Purdy, P.-L. Yu, R.~W. Peterson, N.~S. Kampel, and C.~A. Regal,
  \enquote{Strong {Optomechanical} {Squeezing} of {Light},}
  {\protect\JournalTitle{Physical Review X}} \textbf{3}, 031012 (2013).

\bibitem{nielsen_multimode_2017}
W.~H.~P. Nielsen, Y.~Tsaturyan, C.~B. Møller, E.~S. Polzik, and A.~Schliesser,
  \enquote{Multimode optomechanical system in the quantum regime,}
  {\protect\JournalTitle{Proceedings of the National Academy of Sciences}}
  \textbf{114}, 62--66 (2017).

\bibitem{rossi_measurement-based_2018}
M.~Rossi, D.~Mason, J.~Chen, Y.~Tsaturyan, and A.~Schliesser,
  \enquote{Measurement-based quantum control of mechanical motion,}
  {\protect\JournalTitle{Nature}} \textbf{563}, 53--58 (2018).

\bibitem{brubaker_optomechanical_2022}
B.~Brubaker, J.~Kindem, M.~Urmey, S.~Mittal, R.~Delaney, P.~Burns, M.~Vissers,
  K.~Lehnert, and C.~Regal, \enquote{Optomechanical {Ground}-{State} {Cooling}
  in a {Continuous} and {Efficient} {Electro}-{Optic} {Transducer},}
  {\protect\JournalTitle{Physical Review X}} \textbf{12}, 021062 (2022).

\bibitem{tsaturyan_ultracoherent_2017}
Y.~Tsaturyan, A.~Barg, E.~S. Polzik, and A.~Schliesser, \enquote{Ultracoherent
  nanomechanical resonators via soft clamping and dissipation dilution,}
  {\protect\JournalTitle{Nature Nanotechnology}} \textbf{12}, 776--783 (2017).

\bibitem{mason_continuous_2019}
D.~Mason, J.~Chen, M.~Rossi, Y.~Tsaturyan, and A.~Schliesser,
  \enquote{Continuous force and displacement measurement below the standard
  quantum limit,} {\protect\JournalTitle{Nature Physics}} \textbf{15}, 745--749
  (2019).

\bibitem{chen_entanglement_2020}
J.~Chen, M.~Rossi, D.~Mason, and A.~Schliesser, \enquote{Entanglement of
  propagating optical modes via a mechanical interface,}
  {\protect\JournalTitle{Nature Communications}} \textbf{11}, 943 (2020).

\bibitem{thomas_entanglement_2021}
R.~A. Thomas, M.~Parniak, C.~Østfeldt, C.~B. Møller, C.~Bærentsen,
  Y.~Tsaturyan, A.~Schliesser, J.~Appel, E.~Zeuthen, and E.~S. Polzik,
  \enquote{Entanglement between distant macroscopic mechanical and spin
  systems,} {\protect\JournalTitle{Nature Physics}} \textbf{17}, 228--233
  (2021).

\bibitem{seis_ground_2022}
Y.~Seis, T.~Capelle, E.~Langman, S.~Saarinen, E.~Planz, and A.~Schliesser,
  \enquote{Ground state cooling of an ultracoherent electromechanical system,}
  {\protect\JournalTitle{Nature Communications}} \textbf{13}, 1507 (2022).

\bibitem{kleckner_optomechanical_2011}
D.~Kleckner, B.~Pepper, E.~Jeffrey, P.~Sonin, S.~M. Thon, and D.~Bouwmeester,
  \enquote{Optomechanical trampoline resonators,} {\protect\JournalTitle{Optics
  Express}} \textbf{19}, 19708--19716 (2011).

\bibitem{fan_analysis_2002}
S.~Fan and J.~D. Joannopoulos, \enquote{Analysis of guided resonances in
  photonic crystal slabs,} {\protect\JournalTitle{Physical Review B}}
  \textbf{65}, 235112 (2002).

\bibitem{peng_experimental_1996}
S.~Peng and G.~M. Morris, \enquote{Experimental demonstration of resonant
  anomalies in diffraction from two-dimensional gratings,}
  {\protect\JournalTitle{Optics Letters}} \textbf{21}, 549 (1996).

\bibitem{kanskar_observation_1997}
M.~Kanskar, P.~Paddon, V.~Pacradouni, R.~Morin, A.~Busch, J.~F. Young, S.~R.
  Johnson, J.~MacKenzie, and T.~Tiedje, \enquote{Observation of leaky slab
  modes in an air-bridged semiconductor waveguide with a two-dimensional
  photonic lattice,} {\protect\JournalTitle{Applied Physics Letters}}
  \textbf{70}, 1438--1440 (1997).

\bibitem{suh_displacement-sensitive_2003}
W.~Suh, M.~F. Yanik, O.~Solgaard, and S.~Fan, \enquote{Displacement-sensitive
  photonic crystal structures based on guided resonance in photonic crystal
  slabs,} {\protect\JournalTitle{Applied Physics Letters}} \textbf{82},
  1999--2001 (2003).

\bibitem{lousse_angular_2004}
V.~Lousse, W.~Suh, O.~Kilic, S.~Kim, O.~Solgaard, and S.~Fan, \enquote{Angular
  and polarization properties of a photonic crystal slab mirror,}
  {\protect\JournalTitle{Optics Express}} \textbf{12}, 1575--1582 (2004).

\bibitem{kemiktarak_mechanically_2012}
U.~Kemiktarak, M.~Metcalfe, M.~Durand, and J.~Lawall, \enquote{Mechanically
  compliant grating reflectors for optomechanics,}
  {\protect\JournalTitle{Applied Physics Letters}} \textbf{100}, 061124 (2012).

\bibitem{bui_high-reflectivity_2012}
C.~H. Bui, J.~Zheng, S.~W. Hoch, L.~Y.~T. Lee, J.~G.~E. Harris, and
  C.~Wei~Wong, \enquote{High-reflectivity, high-{Q} micromechanical membranes
  via guided resonances for enhanced optomechanical coupling,}
  {\protect\JournalTitle{Applied Physics Letters}} \textbf{100}, 021110 (2012).

\bibitem{kemiktarak_cavity_2012}
U.~Kemiktarak, M.~Durand, M.~Metcalfe, and J.~Lawall, \enquote{Cavity
  optomechanics with sub-wavelength grating mirrors,}
  {\protect\JournalTitle{New Journal of Physics}} \textbf{14}, 125010 (2012).

\bibitem{makles_2d_2015}
K.~Makles, T.~Antoni, A.~G. Kuhn, S.~Deléglise, T.~Briant, P.-F. Cohadon,
  R.~Braive, G.~Beaudoin, L.~Pinard, C.~Michel, V.~Dolique, R.~Flaminio,
  G.~Cagnoli, I.~Robert-Philip, and A.~Heidmann, \enquote{{2D} photonic-crystal
  optomechanical nanoresonator,} {\protect\JournalTitle{Optics Letters}}
  \textbf{40}, 174--177 (2015).

\bibitem{stambaugh_membrane---middle_2015}
C.~Stambaugh, H.~Xu, U.~Kemiktarak, J.~Taylor, and J.~Lawall, \enquote{From
  membrane-in-the-middle to mirror-in-the-middle with a high-reflectivity
  sub-wavelength grating,} {\protect\JournalTitle{Annalen der Physik}}
  \textbf{527}, 81--88 (2015).

\bibitem{norte_mechanical_2016}
R.~Norte, J.~Moura, and S.~Gröblacher, \enquote{Mechanical {Resonators} for
  {Quantum} {Optomechanics} {Experiments} at {Room} {Temperature},}
  {\protect\JournalTitle{Physical Review Letters}} \textbf{116}, 147202 (2016).

\bibitem{chen_high-finesse_2017}
X.~Chen, C.~Chardin, K.~Makles, C.~Caër, S.~Chua, R.~Braive, I.~Robert-Philip,
  T.~Briant, P.-F. Cohadon, A.~Heidmann, T.~Jacqmin, and S.~Deléglise,
  \enquote{High-finesse {Fabry}–{Perot} cavities with bidimensional {Si3N4}
  photonic-crystal slabs,} {\protect\JournalTitle{Light: Science \&
  Applications}} \textbf{6}, e16190--e16190 (2017).

\bibitem{zhou_cavity_2022}
F.~Zhou, Y.~Bao, J.~J. Gorman, and J.~Lawall, \enquote{Cavity optomechanical
  bistability with an ultrahigh reflectivity photonic crystal membrane,}
  (2022). ArXiv:2211.10485 [physics].

\bibitem{villanueva_evidence_2014}
L.~Villanueva and S.~Schmid, \enquote{Evidence of {Surface} {Loss} as
  {Ubiquitous} {Limiting} {Damping} {Mechanism} in {SiN} {Micro}- and
  {Nanomechanical} {Resonators},} {\protect\JournalTitle{Physical Review
  Letters}} \textbf{113}, 227201 (2014).

\bibitem{moura_centimeter-scale_2018}
J.~P. Moura, R.~A. Norte, J.~Guo, C.~Schäfermeier, and S.~Gröblacher,
  \enquote{Centimeter-scale suspended photonic crystal mirrors,}
  {\protect\JournalTitle{Optics Express}} \textbf{26}, 1895--1909 (2018).

\bibitem{fox_resonant_1961}
A.~G. Fox and T.~Li, \enquote{Resonant {Modes} in a {Maser} {Interferometer},}
  {\protect\JournalTitle{The Bell System Technical Journal}} p.~36 (1961).

\bibitem{asoubar_simulation_2016}
D.~Asoubar, \enquote{Simulation of {Continuous}-{Wave} {Solid}-{State} {Laser}
  {Resonators} using {Field} {Tracing} and a {Fully} {Vectorial} {Fox}-{Li}
  {Algorithm},} Ph.D. thesis, Friedrich-Schiller University Jena, Jena (2016).

\bibitem{dorsel_optical_1983}
A.~Dorsel, J.~D. McCullen, P.~Meystre, E.~Vignes, and H.~Walther,
  \enquote{Optical {Bistability} and {Mirror} {Confinement} {Induced} by
  {Radiation} {Pressure},} {\protect\JournalTitle{Physical Review Letters}}
  \textbf{51}, 1550--1553 (1983).

\bibitem{dumont_flexure-tuned_2019}
V.~Dumont, S.~Bernard, C.~Reinhardt, A.~Kato, M.~Ruf, and J.~C. Sankey,
  \enquote{Flexure-tuned membrane-at-the-edge optomechanical system,}
  {\protect\JournalTitle{Optics Express}} \textbf{27}, 25731 (2019).

\bibitem{saarinen_laser_2022}
S.~A. Saarinen, N.~Kralj, E.~C. Langman, Y.~Tsaturyan, and A.~Schliesser,
  \enquote{Laser cooling a membrane-in-the-middle system close to the quantum
  ground state from room temperature,}  (2022). ArXiv:2206.11169 [physics,
  physics:quant-ph].

\bibitem{guo_integrated_2017}
J.~Guo, R.~A. Norte, and S.~Gröblacher, \enquote{Integrated optical force
  sensors using focusing photonic crystal arrays,}
  {\protect\JournalTitle{Optics Express}} \textbf{25}, 9196 (2017).

\bibitem{borkje_critical_2020}
K.~Børkje, \enquote{Critical quantum fluctuations and photon antibunching in
  optomechanical systems with large single-photon cooperativity,}
  {\protect\JournalTitle{Physical Review A}} \textbf{101}, 053833 (2020).

\bibitem{meng_mechanical_2020}
C.~Meng, G.~A. Brawley, J.~S. Bennett, M.~R. Vanner, and W.~P. Bowen,
  \enquote{Mechanical {Squeezing} via {Fast} {Continuous} {Measurement},}
  {\protect\JournalTitle{Physical Review Letters}} \textbf{125}, 043604 (2020).

\end{thebibliography}

\end{document}